\documentclass{emulateapj}

\usepackage{natbib}

\newcommand{\nar}{New Astronomy Reviews}
\defcitealias{Doi:2013a}{D13}

\slugcomment{Submitted October 17, 2014, Accepted December 9, 2014}

\shorttitle{FR-I candidate in the NLS1 Mrk~1239}
\shortauthors{Doi, et al.}

\begin{document}

\title{A Fanaroff-Riley Type~I Candidate in Narrow-Line Seyfert 1 Galaxy Mrk~1239}

\author{Akihiro Doi\altaffilmark{1,2}, Kiyoaki Wajima\altaffilmark{3}, Yoshiaki Hagiwara\altaffilmark{4}, and Makoto Inoue\altaffilmark{5}}

\altaffiltext{1}{The Institute of Space and Astronautical Science, Japan Aerospace Exploration Agency, 3-1-1 Yoshinodai, Chuou-ku, Sagamihara, Kanagawa 252-5210, Japan}\email{akihiro.doi@vsop.isas.jaxa.jp}
\altaffiltext{2}{Department of Space and Astronautical Science, The Graduate University for Advanced Studies, 3-1-1 Yoshinodai, Chuou-ku, Sagamihara, Kanagawa 252-5210, Japan}
\altaffiltext{3}{Korea Astronomy and Space Science Institute, 776 Daedeokdae-ro, Yuseong, Daejeon 305-348, Korea}
\altaffiltext{4}{National Astronomical Observatory of Japan, 2-21-1 Osawa, Mitaka, Tokyo 181-8588, Japan}
\altaffiltext{5}{Academia Sinica Institute of Astronomy and Astrophysics, P.O. Box 23-141, Taipei 10617, Taiwan}

\begin{abstract}
We report finding kiloparsec-scale radio emissions aligned with parsec-scale jet structures in the narrow-line Seyfert~1~(NLS1) galaxy Mrk~1239 using the Very Large Array and the Very Long Baseline Array.  Thus, this radio-quiet NLS1 has a jet-producing central engine driven by essentially the same mechanism as that of other radio-loud active galactic nuclei~(AGNs).  Most of the radio luminosity is concentrated within 100~parsecs and overall radio morphology looks edge-darkened; the estimated jet kinetic power is comparable to Fanaroff--Riley Type~I radio galaxies.  
The conversion from accretion to jet power appears to be highly inefficient in this highly accreting low-mass black hole system compared with that in a low-luminosity AGN with similar radio power driven by a sub-Eddington, high-mass black hole.  Thus, Mrk~1239 is a crucial probe to the unexplored parameter spaces of central engines for a jet formation.  
\end{abstract}


\section{INTRODUCTION}\label{section:introduction} 
Narrow-line Seyfert~1~(NLS1) galaxies, a subclass of active galactic nuclei~(AGNs), are generally observed as weak radio sources.  NLS1s are identified by their optical properties \citep{Osterbrock:1985, Goodrich:1989}.  The currently used classification \citep{Pogge:2000} includes (1)~narrow permitted lines only slightly broader than forbidden lines, (2)~full width at half maximum~(FWHM; H$\beta)<2000$~km~s$^{-1}$, and (3)~flux ratio [\ion{O}{3}]/H$\beta<3$, but exceptions are allowed if there is strong [\ion{Fe}{7}] and [\ion{Fe}{5}] present, unlike that observed in type-2 Seyfert galaxies.  Many other peculiarities are also observed, such as strong permitted \ion{Fe}{2} emission lines \citep[e.g.][]{Boroson:1992}, rapid X-ray variability \citep[e.g.][]{Leighly:1999a}, and prominent soft X-ray excess \citep[e.g.][]{Boller:1996}.  These extremes are presumably relevant to high mass accretion rates close to the Eddington limit \citep[e.g.][]{Boroson:1992} on relatively low mass black holes \citep[$\sim 10^5$--10$^{7.5}$~M$_\sun$; e.g., ][]{Zhou:2006}.  On the fundamental plane of black hole activity \citep{Merloni:2003}, highly accreting and small-mass black hole systems, such as NLS1s, tend to be radio-quiet\footnote{Radio loudness $R$ is defined as the ratio of the 5-GHz radio and optical {\it B}-band flux densities.  The threshold of $R=10$ separating radio-loud and radio-quiet objects \citep{Kellermann:1989} is frequently used (however, see \citealt{Ho:2001a}).}.    
In fact, the observed fraction of radio-loud objects in the NLS1 population is significantly low compared with that in normal Seyfert galaxies and quasars \citep[e.g.,][]{Greene:2006,Zhou:2006}.  Thus, radio jets are apparently powered less efficiently by black holes of lower mass at higher accretion rates.   

Kiloparsec-scale radio emissions have been exceptionally detected in several radio-loud NLS1s thus far \citep{Anton:2008,Gliozzi:2010,Doi:2012}; the detection rate is lower than that in broad-line AGNs in a statistical sense \citep{Doi:2012}.  Moreover, $\gamma$-ray emissions have been detected in several NLS1s \citep[e.g.,][]{Abdo:2009}.  The existence of these exceptions indicates that NLS1 central engines also have the ability to be a part of blazars and radio galaxies as radio sources.  Very-long-baseline interferometry (VLBI) observations have provided evidence for nonthermal jets with very high brightness temperatures in at least several radio-loud NLS1s \citep{Doi:2006,Doi:2007,Doi:2011a,Giroletti:2011,DAmmando:2012,DAmmando:2013a,Wajima:2014}.  Furthermore, \citet{Doi:2013a}, hereafter \citetalias{Doi:2013a}, have reported VLBI detections of five of seven radio-quiet NLS1s, also implying nonthermal jets (see also \citealt{Giroletti:2005} for NGC~4051; \citealt{Middelberg:2004} for NGC~5506).  These observations suggest that the radio emission process in both radio-quiet and radio-loud NLS1s is essentially the same as that observed in other radio-loud AGN classes.  Thus, NLS1 radio sources potentially provide irreplaceable clues to jet phenomena at the extreme end of the parameter space for black hole activities.  However, little information is available for detailed properties of the radio-quiet subclass in particular.

Mrk~1239 is an NLS1 as observed with FWHM(H$\beta)=1075$~km~s$^{-1}$, [\ion{O}{3}]/H$\beta=1.29$, and [\ion{Fe}{2}]/H$\beta=0.63$ \citep{Veron-Cetty:2001a}.  
Black hole mass has been estimated \citep{Ryan:2007} to be  
$M_\mathrm{BH}=7.8 \times 10^5\ M_\sun$ on the basis of the FWHM(H$\beta$)--$L_{\lambda5100}$ relation \citep{Kaspi:2005}, and 
$M_\mathrm{BH}=1.3 \times 10^6\ M_\sun$ on the basis of the FWHM(H$\beta$)--$L_{H\beta}$ relation \citep{Greene:2005}.      
The Hubble type of the host is E--S0 (the NASA/IPAC Extragalactic Database: NED).  
\citet{Ryan:2007} analyzed the surface profile of the spheroidal component in Mrk~1239, which shows a S$\mathrm{\acute{e}}$rsic index of $n=1.65$ (1.08) and an effective radius of only 0.17~kpc (0.19~kpc) at $J$-band ($K_\mathrm{s}$-band).  The result suggests an estimated black hole mass of $M_\mathrm{BH} = 6.5 \times 10^6\ M_\sun$ ($5.3 \times 10^5\ M_\sun$) in a pseudobulge, according to the $n$--$M_\mathrm{BH}$ relation \citep{Graham:2007}.   
Thus, Mrk~1239 is virtually assured to have a small mass black hole.  
The Eddington ratio is quite high ($\sim2$--3) and the soft X-ray spectrum is unusually steep with a spectral index of $\sim3$ \citep{Grupe:2004a}, which indicate a genuine NLS1.     
Mrk~1239 is a weak radio source in mJy level as a radio-quiet AGN with $R=0.54$\footnote{Using the Sloan Digital Sky Survey~(SDSS) Data Release~7 (DR7) PSF magnitudes of 16.237~mag and 15.292~mag at $u$- and $g$-bands, respectively, we derive a $B$-band flux density of 40.1~mJy, which is corrected in terms of reddening ($A_\mathrm{B}=3.0$~mag estimated from H$\alpha$/H$\beta$=6.4 \citep{Grupe:1998} assuming intrinsic Balmer decrement of $3.1$, and the standard extinction law of \citealt{Cardelli:1989}) and $k$-correction assuming a spectral index of $-0.61$ on average for the NLS1 sample \citep{Grupe:2010}.  Radio flux density at 5~GHz is 22.7~mJy, which is derived from observed VLA flux densities (Section~\ref{section:observation}) and $k$-correction assuming a spectral index of $-0.84$ (Section~\ref{section:observation}).  \citet{Grupe:2004a} derived $R=5.3$--7.0 based on their optical observation; the optical nucleus may be variable.  This value also indicates a radio-quiet source; however, it is near the borderline between radio-loud and radio-quiet.  We cannot rule out that the radio loudness becomes larger, if nuclear extraction is properly performed \citep{Ho:2001a}; however, the optical nucleus in a Hubble Space Telescope image was saturated \citep{Capetti:2007}.}.      
A Very Large Array~(VLA) observation at 8.5~GHz revealed a slightly elongated structure with $\sim0\farcs12$ ($\sim50$~pc; \citetalias{Doi:2013a}), while earlier studies reported that it is apparently compact \citep{Orienti:2010,Thean:2000,Ulvestad:1995}.      
Mrk~1239 is one of a few nearby radio-quiet NLS1s with a pc-scale structure significantly resolved by the VLBI technique at a milli-arcsecond~(mas) resolution \citepalias{Doi:2013a}.  However, the structure was resolved into only an ambiguous feature consisting of several components under a limited image dynamic range (\citetalias{Doi:2013a}, \citealt{Orienti:2010}).  

In the present paper, we report new VLBI and VLA images of Mrk~1239 with significantly improved qualities, which allow us to derive detailed jet properties on this highly accreting and small-mass black hole system.  In Section~\ref{section:observation}, radio observations and data reduction are described, and the results are presented in Section~\ref{section:results}.  In Section~\ref{section:discussion}, we provide short discussions.  
A cosmology with $H_0=70.5$~km~s$^{-1}$~Mpc$^{-1}$, $\Omega_\mathrm{M}=0.27$, $\Omega_\mathrm{\Lambda}=0.73$ is adopted.  The redshift of Mrk~1239 is $z=0.019927 \pm 0.000127$ \citep{Beers:1995}.  The luminosity distance is 91.2~Mpc and the angular-size distance is 87.4~Mpc; 1~mas corresponds to 0.424~pc at the distance to Mrk~1239.

\section{Observations and Data Reductions}\label{section:observation}
\begin{figure*}
\epsscale{1.0}
\plotone{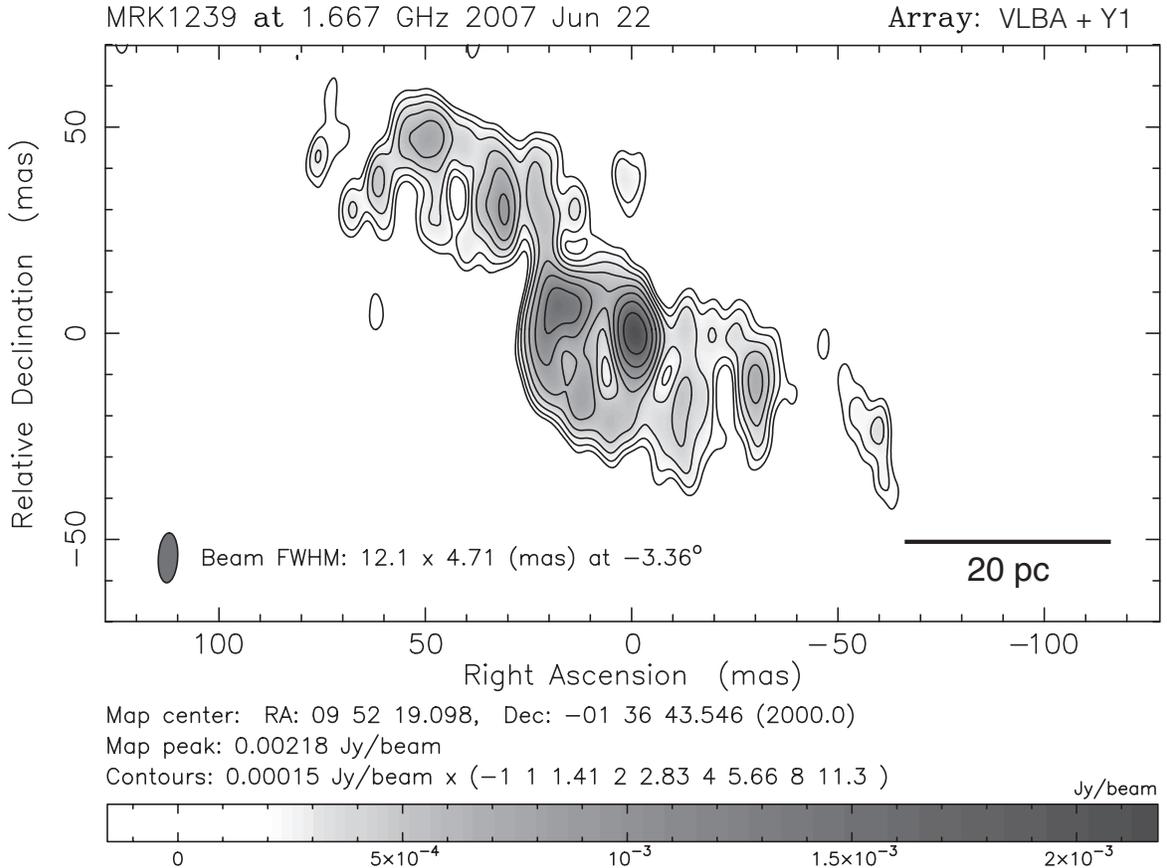}
\figcaption{VLBI image of Mrk~1239 obtained using VLBA+Y1 at 1.7~GHz.  Contour levels are separated by factors of $\sqrt{2}$ beginning at $3\sigma$ of the r.m.s.~image noise of $\sigma = 50$~$\mu$Jy~beam$^{-1}$.  Beam size is illustrated as a gray ellipse at the lower left.\label{figure1}}
\end{figure*}

  The VLBI observation of Mrk~1239 was performed using 10~antennas of the Very Long Baseline Array~(VLBA) plus one antenna of the VLA~(``VLBA+Y1'') on June~22 and 23, 2007, for 6~hours each day, using phase-reference mode (project code: BD124).  Dual circular polarizations were obtained at a center frequency of 1.667~GHz ($\lambda18.0$~cm) with a total bandwidth of 256~MHz.  Data reduction was performed using the Astronomical Image Processing System~(AIPS) according to the standard procedures for VLBA phase-referencing.  The solutions of self-calibration on a calibrator source in both amplitude and phase determined by using a structure model were applied to the target (Mrk~1239).  Final calibrations on the target data were performed through a few iterations of deconvolution and self-calibration (only in phase).  Using Difmap software, we deconvolved the dirty image to make a create final image using uniform weighting, natural weighting, and uv-tapering step-by-step, and we displayed the final image with natural weighting (Figure~\ref{figure1}). 　The supplement of one VLA antenna to the VLBA contributes to significantly improve sensitivity for lower brightness with baselines shorter than those of the previous VLBA observation (\citetalias{Doi:2013a}, \citealt{Orienti:2010}); the correlated flux density retrieved at the shortest baseline is 45~mJy at $\sim0.05\ \mathrm{M\lambda}$ in the present study (Figure~\ref{figure2}, panel a), while 16~mJy at $\sim1\ \mathrm{M\lambda}$ in the previous study (figure~3, panel a in \citetalias{Doi:2013a}).

  We retrieved two sets of archival data obtained using the VLA A-array configuration~(AM384 and AS633).  AM384 observed dual circular polarizations on January~16, 1993, at a central frequency of 8.465~GHz with a total bandwidth of 100~MHz.  AS633 observed a right circular polarization on April~14, 1998, at the L-band with a total bandwidth of~6.250 MHz; however, we used only the second intermediate frequency (IF2) at a central frequency of 1.581~GHz with a bandwidth 3.125~MHz because of significant radio-frequency interference in IF1.  Data reduction was performed using the AIPS according to the standard procedures for VLA continuum data.  Self-calibration was performed on the targets.  Using Difmap, we create final images (Figure~\ref{figure2}, panels b and c) in the same manner as the VLBI image.  The parameters of the radio images are shown in Table~\ref{table1}.

\section{RESULTS}\label{section:results}
\begin{figure*}
\epsscale{1.15}
\plotone{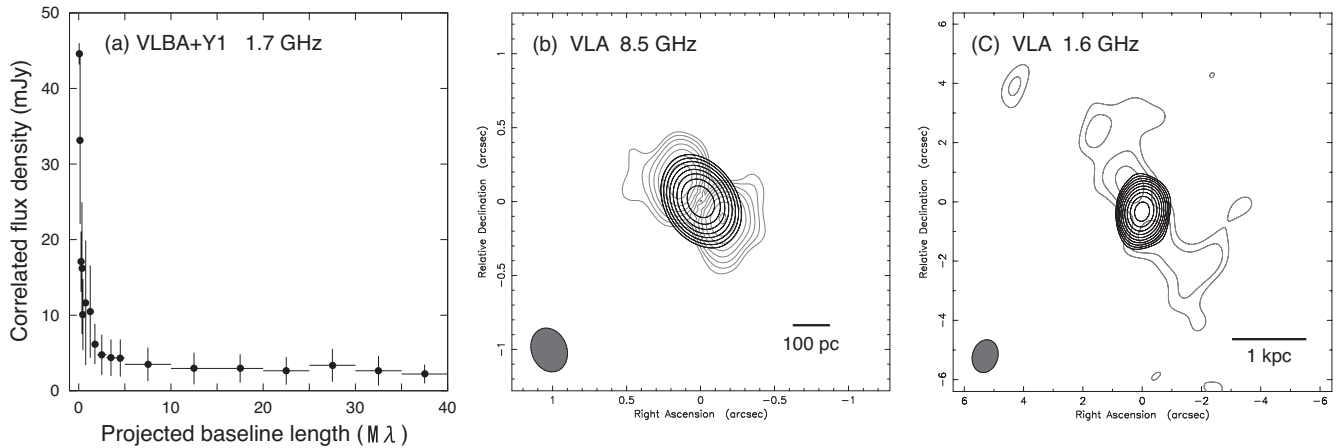}
\figcaption{(a)~Correlated visibility amplitude profile with projected $uv$-distance at $PA=45\degr$ of VLBA+Y1 data for Mrk~1239.  (b)~VLA images of Mrk~1239 with A-array configuration at 8.5~GHz.  Black contours represent intensity map and are separated by factors of $\sqrt{2}$ beginning at $8\sigma$ of the r.m.s.~image noise of $\sigma = 35$~$\mu$Jy~beam$^{-1}$.  Gray contours represent a point-source-subtracted (13.2~mJy) map and are separated by factors of $\sqrt{2}$ beginning at $\sigma/\sqrt{2}$.  Beam size is illustrated as a gray ellipse at the lower left.  (c)~VLA images of Mrk~1239 with A-array configuration at 1.6~GHz.  Black contours are beginning at $6\sigma$ ($\sigma =$338~$\mu$Jy~beam$^{-1}$).  Gray contours represent a point-source-subtracted (56~mJy) map.\label{figure2}}   
\end{figure*}

The VLBA image shows an intensity peak of 2.2~mJy~beam$^{-1}$, corresponding to $T_\mathrm{B} = 2.5 \times 10^7$~K.  
The pc-scale structure is dominated by diffuse emissions with brightness temperatures of $T_\mathrm{B} > 1.7 \times 10^6$~K for surface brightness more than $3\sigma$ of image noise and is elongated to a position angle of 47\degr.
The visibility amplitude profile showed prominent enhancement at short baselines, one-half the correlated flux at $\sim0.22$~M$\lambda$, which corresponds to 430~mas ($\sim200$~pc on two side) and implies $\sim2\times10^5$~K (Figure~\ref{figure2}, panel a).  
The position angle of the radio morphology is not aligned along the major axis of the host galaxy in an optical image ($PA=153.4\degr$ in SDSS).  In addition, the standard FIR/radio ratio $q$ \citep[][and references therein]{Condon:1992} is 1.6 or less for Mrk~1239\footnote{This is calculated from {\it IRAS} flux densities of 1.3~Jy and $<2.4$~Jy at $60\ \mu$m and $100\ \mu$m, respectively, and VLBA total flux density (Table~\ref{table1}) and the radio spectral index $\alpha=-0.84$.}, which is significantly smaller than $q\sim2.3 \pm 0.2$ for starburst galaxies.  Thus, the stellar origin cannot be responsible for the radio emissions detected using the VLBA.  These are evidence of nonthermal synchrotron emissions as AGN jet activity.

\begin{table*}
\caption{Parameters of radio images\label{table1}}
\begin{center}
\begin{footnotesize}
\begin{tabular}{llccccccc}
\hline
\hline
Telescope&	Obs.~date&	$\nu$&	$I_\nu$&	$S_\nu$&	$\sigma$&	$\theta_\mathrm{maj}$&	$\theta_\mathrm{min}$&	$P.A.$ \\
&	&	(GHz)&	(mJy beam$^{-1}$)&	(mJy)&	(mJy beam$^{-1}$)&	(\arcsec)&	(\arcsec)&	(\degr) \\
(1)&	(2)&	(3)&	(4)&	(5)&	(6)&	(7)&	(8)&	(9) \\
\hline
VLBA+Y1&	June 22--23, 2007& 	1.667&	$2.2\pm0.1$&	$41.8\pm2.1$&	0.050&	0.0121&	0.0047&	$-3.4$ \\
VLA-A&	January 16, 1993&	8.465&	$13.3\pm0.7$&	$14.6\pm0.7$&	0.035&	0.30&	0.24&	$+21$ \\
VLA-A&	April 14, 1998&	1.581&	$57.9\pm2.9$ &	$59.8\pm3.0$ &	0.338 &	1.13&	0.86&	$-17$ \\
\hline
\end{tabular}
\tablecomments{Col.~(1) telescope; Col.~(2) observation date; Col.~(3) observing frequency; Col.~(4) peak intensity; Col.~(5) total flux density; Col.~(6) image r.m.s.~noise; Col.~(7)--(9) major and minor axes and position angle of synthesized beam, respectively.}
\end{footnotesize}
\end{center}
\end{table*}

The pc-scale structure shows jet-like morphology, which is apparently two-sided with respect to the intensity peak, with a one-side extent of $\sim80$~mas, corresponding to $\sim34$~pc and $\sim 3 \times 10^8 R_\mathrm{S}$ assuming $M_\mathrm{BH} \approx 1 \times 10^6 M_\sun$ (see, Section~\ref{section:introduction} and references therein), where $R_\mathrm{S}$ is the Schwarzschild radius.  The morphology is quite different from that of regular radio-loud AGNs showing a prominent core + a one-sided collimated jet; Mrk~1239's jet is not so strongly beamed.  It is not clear which jets are approaching or receding.  The transverse structure of jets is significantly resolved on both sides; the deconvolved width is approximately 20~mas, corresponding to $\sim 8$~pc and $\sim 9 \times 10^7 R_\mathrm{S}$.    
Thus, the large width of jets is a noteworthy characteristics discovered from the VLBA imaging.       
The opening angle of jets cannot be determined because the location of core is unknown.

The VLA images are dominated by an apparently unresolved component; however, point-source-subtracted maps clearly reveal the presence of extended emissions on both sides at a similar position angle to the parsec scales.  The lengths on one side are $\sim0\farcs2$ and $\sim3\farcs3$ corresponding to $\sim85$~pc and $\sim1.4$~kpc at 8.5~GHz and 1.6~GHz, respectively.  The discovery of the kpc-scale radio structure is a significant conclusion from the VLA images.  The maximum extent of $\sim1.4$~kpc corresponds to $\sim 1 \times 10^{10} R_\mathrm{S}$ for Mrk~1239.    

Extended components in the arcsecond resolutions contribute a maximum of only $\sim10$\% in total radio emissions.  The radio morphology is highly concentrated on the central region of $<100$~pc.

\section{DISCUSSION}\label{section:discussion}
Radio imaging revealed nonthermal jets showing not well-collimated two-sided morphology with relatively high brightness temperatures at pc scales and extending up to $\sim1.4$~kpc with a low brightness.  Most radiated power is concentrated within $<100$~pc.  Thus, the radio profile is so called ``edge-darkened'' as a whole, resembling Fanaroff-Riley Type~I \citep{Fanaroff:1974}.  However, Mrk~1239 radio source is significantly smaller than general radio galaxies in the physical scale.  Nevertheless, the radio structure extends to $\sim 1 \times 10^{10} R_\mathrm{S}$, which is comparable, in the unit of $R_\mathrm{S}$, to the size of largest ($\sim1$~Mpc) radio galaxies with $M_\mathrm{BH} \approx 1 \times 10^9 M_\sun$.    

The edge-darkened morphology may reflect the subsonic speed of advancing radio lobes.  As one of the many suggestions in the literature trying to explain the FR-I/FR-II dichotomy for general radio galaxies, the FR-I/FR-II dichotomy is determined by the ratio of the jet kinetic power $L_\mathrm{j}$ to the ambient number density $\bar{n}_\mathrm{a}$ at the core radius ($\sim1$~kpc) of the host galaxy \citep[e.g.,][]{Kaiser:2007};  
a threshold exceeding $L_\mathrm{j} /\bar{n}_\mathrm{a}=10^{44}$--$10^{45}$~ergs~s$^{-1}$~cm$^{-3}$ is required to extend supersonic lobes beyond a core radius without disrupting hot spots for a young radio galaxy \citep{Kawakatu:2008}.   
The jet kinetic power of Mrk~1239 is expected to be only $\sim10^{43.0}$~ergs~s$^{-1}$, if we use the empirical relation between the cavity power as a proxy for mechanical jet power and 1.4-GHz radio power for general radio galaxies \citep{OSullivan:2011}.  
Furthermore, a dense environment can be inferred from the relatively large dust extinction of H$\alpha$/H$\beta$=6.4 \citep{Grupe:1998}, implying $A_\mathrm{V}=$2.3~mag, and a natural column density of $N_\mathrm{H} \sim 3 \times 10^{23}$~cm$^{-2}$ \citep{Grupe:2004a} through the nucleus of Mrk~1239.  Thus, the putative hot spots could be disrupted immediately after their launch due to insufficient jet kinetic power from this small-mass black hole system.

FR~I-like morphology in a weak radio source recalls the low-luminosity AGN NGC~4278, which is placed at the low-power end of the correlation between optical and radio core luminosity in FR~I radio galaxies \citep{Capetti:2002}.  The VLBI images of NGC~4278 show two-sided pc-scale morphology with significantly resolved jets with a similar transverse width ($\sim10$~pc, \citealt{Giovannini:2001,Giroletti:2005}) and an off-core emission of extent only $\sim1\arcsec$ ($\sim0.1$~kpc)  accounting for $\sim10$\% in total emission \citep{Wrobel:1984}.  Thus, it resembles Mrk~1239 in terms of radio morphology.  Their radio luminosities (relevant to jet kinetic power) are also similar: $10^{38.9}$~ergs~s$^{-1}$ and $10^{38.4}$~ergs~s$^{-1}$ at 1.4~GHz for Mrk~1239 and NGC~4278, respectively.  
On the contrary, their accretion powers differ significantly, because of 2--10~keV X-ray luminosities of $L_\mathrm{X} = 2.5\times 10^{42}$~ergs~s$^{-1}$ and $9.1 \times 10^{39}$~ergs~s$^{-1}$ for Mrk~1239 and NGC~4278, respectively\footnote{Distance and energy range have been corrected from the original values of \citet{Grupe:2004a} and \citet{Younes:2010}.}.  Thus, the central engine of Mrk~1239 has significantly lower efficiency in jet generation.  

The difference between the two central engines are potentially responsible for quite different conversion efficiencies from the accretion power to the jet power because their black hole mass and accretion rate are at opposite sides: $M_\mathrm{BH} \approx 1 \times 10^6 M_\sun$ (see, Section~\ref{section:introduction} and references therein) and $L_\mathrm{X}/L_\mathrm{Edd} \approx 0.02$ for Mrk~1239 and $M_\mathrm{BH} = 3.7 \times 10^8 M_\sun$ and $L_\mathrm{X}/L_\mathrm{Edd} = 2\times10^{-7}$ for NGC~4278.          
The tendency of jet conversion becoming less efficient at higher accretion rates has been previously suggested by \citet{Merloni:2007} according to their sample of radio galaxies, which includes objects with jet powers similar to Mrk~1239 but sub-Eddington accretion onto black holes with masses ranging from $10^{8.2}\ M_\sun$ to $10^{9.5}\ M_\sun$: $\log{L_\mathrm{j}/L_\mathrm{Edd}}=0.49 \log{5L_\mathrm{X}/L_\mathrm{Edd}}-0.78$.  The nature in high Eddington ratios and lower mass ranges was unclear thus far.  It is noteworthy that Mrk~1239 smoothly follows the trend as higher Eddington ratio extrapolation.   Thus, Mrk~1239 is crucial for probing the unexplored parameter spaces of black hole mass and accretion rate for jet conversion efficiency of central engines.  We will discuss the jet efficiency in an entire range of parameter spaces, including NLS1s, in a future paper.  

By deep imaging with a specially-arranged VLBI with short baselines (VLBA+Y1), we have discovered FR~I-like morphology in one of radio-quiet sources with NLS1 nuclei, while the previous images (\citetalias{Doi:2013a}, \citealt{Orienti:2010}) showed only the slightest sign of structure in a limited sensitivity in brightness.  The case of E1821+643 is also an example of the discovery of a 300~kpc-scale FR-I jets in a radio-quiet quasar by deep imaging \citep{Blundell:2001}.  Both Mrk~1239 and E1821+643 are categorized into radio-quiet AGNs but around a borderline between radio-loud and radio-quiet.  It might be possible that FR~I-like jets are prevalent in radio-intermediate and radio-quiet AGNs essentially; their low-brightness radio structures would be detectable only by deepest imaging with high-sensitivity arrays such as VLBIs involving short baselines, the Karl G.~Jansky Very Large Array, and the future Square Kilometer Array.

\acknowledgments
We are grateful to the anonymous referee for offering constructive comments, which have contributed to substantially improving this paper.  
The VLBA and VLA are operated by the National Radio Astronomy Observatory, which is a facility of the National Science Foundation operated under cooperative agreement by Associated Universities, Inc. 
This study was partially supported by Grants-in-Aid for Scientific Research (B; 24340042, AD) from the Japan Society for the Promotion of Science (JSPS).



\end{document}